\documentclass[12pt,preprint]{aastex}

\newcommand\MSUNYR{\rm M_{\odot}\,yr^{-1}}
\newcommand\MSUN{\rm M_{\odot}}
\newcommand\LSUN{\rm L_{\odot}}
\newcommand\RSUN{\rm R_{\odot}}
\newcommand\Mdot{ \dot{M}}
\newcommand\etal{{\rm et al}. }
\newcommand\be {\begin{equation}}
\newcommand\en{\end{equation}}
\newcommand\cm{\rm cm}

\newcommand\K{\rm K}
\newcommand\coku{CoKu~Tau/4}
\newcommand{\mrm}[1]{\mathrm{#1}}
\newcommand{\wall}{\mathrm{wall}}
\newcommand{\J}{\mathrm{J}}
\renewcommand{\K}{\mathrm{K}}
\renewcommand{\H}{\mathrm{H}}

\def\cm2g{\rm cm^2 \ g^{-1}}

\shorttitle{Truncated Disk of \coku\ }
\shortauthors{D'Alessio \etal}

\begin{document}
 
\title{The Truncated Disk of \coku\ }

\author{
Paola D'Alessio\altaffilmark{1}, 
Lee Hartmann\altaffilmark{2}, Nuria Calvet \altaffilmark{2}, 
Ramiro Franco-Hern\'andez\altaffilmark{1}, 
  William J. Forrest\altaffilmark{3}, Ben Sargent\altaffilmark{3}, 
Elise Furlan\altaffilmark{4}, Keven   
 Uchida\altaffilmark{4}, Joel D. Green\altaffilmark{3}, 
Dan M. Watson\altaffilmark{3}, Christine H. Chen\altaffilmark{5}, 
F. Kemper\altaffilmark{6},  G.C. Sloan\altaffilmark{4}, 
 Joan Najita\altaffilmark{7}
}

\altaffiltext{1}{Centro de Radioastronom\'\i a y Astrof\'\i sica,
Ap.P. 72-3 (Xangari), 58089 Morelia, Michoac\'an, M\'exico}
\altaffiltext{2}{Harvard-Smithsonian Center for Astrophysics, 60 Garden St.,
Cambridge, MA 02138, USA.}
\altaffiltext{3}{Department of Physics and Astronomy, University of Rochester, Rochester, NY 14627-0171}
\altaffiltext{4}{Center for Radiophysics and Space Research, Cornell University, Space Sciences Building, Ithaca, NY 14853-6801}
\altaffiltext{5}{National Research Council Resident Research Associate, Jet Propulsion Laboratory, M/S 169-506, California Institute of Technology, 4800 Oak Grove Drive, Pasadena, CA 91109}
\altaffiltext{6}{Department of Physics and Astronomy, UCLA, 405 Hilgard Avenue, Los Angeles, CA 90095-1562}
\altaffiltext{7}{NOAO, 50 North Cherry Avenue, Tucson, AZ 85719.}

\email{p.dalessio@astrosmo.unam.mx}

\begin{abstract}
We present a  model of a dusty disk with
an inner hole which accounts for the Spitzer Space Telescope Infrared
Spectrograph observations of the low-mass pre-main sequence star
\coku.
We have modeled the mid-IR spectrum (between 8 and 25 $\mu$m) 
as arising from the inner wall  of a disk.  
Our model disk has an evacuated inner zone of radius $\sim$ 10 AU, with a
dusty inner ``wall'', of half-height $\sim 2$~AU, that is 
illuminated at normal incidence by the central star.
The radiative equilibrium temperature decreases from the 
inner disk edge outward through 
the optically-thick disk; this temperature gradient
is responsible for the emission of the silicate bands at 10 and 20 $\mu$m.
The observed spectrum is consistent with being produced by  
Fe-Mg amorphous  glassy olivine and/or pyroxene, with no evidence 
of a crystalline component. 
The mid-infrared spectrum of \coku\  is reminiscent of that of the
much older star TW Hya, where it has been suggested that the 
  significant clearing of its inner disk is due   
  to planet formation.
However, no inner disk remains in \coku, consistent
with the star being a weak-emission (non-accreting) T Tauri star.
The relative
youth of \coku\  ($\sim 1$~Myr) may indicate much more rapid planet
formation than typically assumed.
\end{abstract}
\keywords{planetary systems: formation, protoplanetary disks,  
stars:  pre-main sequence, circumstellar matter}

\section{Introduction}
\label{sec_intro}

The formation of planets around young stars is thought to involve the 
eventual clearing of the natal dusty protoplanetary disks, as planets
accrete and sweep out material.  Coagulation and accretion processes are
expected to occur fastest in inner disk regions, where surface densities
are higher 
and orbital periods shorter
(e.g., Wuchterl, Guillot, \& Lissauer 2000, and references
therein).  An observational signature of inner disk gaps or holes
would be the reduction or elimination of emission from hot dust,
corresponding to a decrease in the short-wavelength infrared emission.

The low-mass pre-main sequence star \coku\  exhibits a spectral
energy distribution (SED) with properties suggesting a substantial
dusty disk with an evacuated inner region; strong infrared excess emission
is detected at wavelengths $\gtrsim 10$ $\mu$m, with little or no
excess over photospheric fluxes
at shorter wavelengths, as shown by observations with the Infrared 
Spectrograph (IRS) on the Spitzer Space Telescope (Forrest \etal 2004).
The interpretation is that \coku\  has a dusty disk which is mostly
cleared of small dust grains inside a radius of 10 AU.
The qualitative nature of the SED of \coku\ 
is reminiscent of that of the Classical T Tauri 
stars TW Hya (Calvet \etal 2002; Uchida \etal 2004), 
GM Aur (Marsh \& Mahoney 1992, 1993; Rice \etal 2003),
and the Herbig Ae star HD 100546 (Bouwman \etal 2003).
Quillen \etal (2004) suggest that the inner disk hole in \coku\ 
might be produced by a relatively low-mass planet; if this is the
case, \coku\  is particularly interesting because its relative  youth
($\sim $ 1 Myr vs. 10 Myr for TW Hya; Webb \etal 1999), 
places severe constraints on 
planetary formation (e.g., Wuchterl \etal 2000, and references therein).

In this paper we present   radiative transfer
models of \coku\  which account for the IRS spectrum and other
constraints on the SED.  We confirm our previous estimate of the
disk inner radius (Forrest \etal 2004), and show that the observed 10 
$\mu$m silicate  
emission feature is likely to be produced in the hot outer layer
of the disk ``wall'' facing the star, without requiring any small
dust inside the inner hole.  We also comment on the dust compositions 
consistent with the observations.

\section{Stellar properties, extinction, and SED}
\label{secc_star}

\coku\  is a weak-emission T Tauri star
(WTTS) of spectral type of M1.5 (Cohen \& Kuhi 1979), with 
no detectable short wavelength (UV) excess and an  
H$\alpha$ equivalent width $\sim$ 1.8 \AA\ (Cohen \& Kuhi 1979).
We assume a distance of 140 pc to the star, based on its location
in the Taurus-Auriga molecular cloud (Kenyon, Dobrzycka, \& Hartmann 1994).

To 
estimate 
the stellar parameters, we need to
quantify the extinction to the star.  Cohen \& Kuhi (1979) found
$A_V = 0.94 \pm 0.29$ from their spectral type determination, adoption
of the standard reddening law, and using fluxes from their scanner spectra
in wavelength regions near 5400, 6000, and 6700 \AA.  However, such a low
extinction is not consistent with relatively recent observations of
near-infrared colors.  Kenyon \& Hartmann (1995; KH95) find $\J - \H = 1.10$,
and the 2MASS catalog yields $\J - \H = 1.08$, while from Table A4 of
KH95, the unreddened color of an M1.5 star should be $\J - \H = 0.67$.
Using an extinction law with $R_V \sim 3.6$, as indicated by
the recent study of Calvet \etal (2004), this would suggest
$A_V \sim 3$.  The $\H-\K$ colors suggest slightly larger extinctions,
but the latter result is much more sensitive to errors in 
photometry and extinction laws than the result obtained from the  $\J - \H$ 
colors.

Recent optical spectra of \coku\  from the FAST spectrograph on the Fred L. Whipple 
1.5m telescope on Mt. Hopkins (discussed in Kenyon \etal 1998) were inspected. 
Although conditions were not photometric,
comparison with other stars observed on the same night indicate a much redder spectrum
than shown by Cohen and Kuhi (1979), roughly consistent with $A_V \sim 3$,
in agreement with the values of extinction inferred from 2MASS colors.
For the purposes of this paper we therefore adopt $A_V = 3$. 
Further optical monitoring of \coku\  would be desirable.

Following KH95, we assign a stellar effective temperature of $T_*=$3720 K 
to \coku.  For $A_V = 3$ we find a stellar luminosity of
approximately $L_*=0.61$ $\LSUN$, with an error of order 20\% or
more given possible variability and the uncertainty in extinction corrections.  
(According to 2MASS, $\J = 10.16$, while KH95 cite 
$\J = 10.37$.) Evolutionary tracks from Siess, Dufour, \& Forestini (2000) 
give a mass 
of $M_*\sim 0.5$ $\MSUN$ and the age of $\sim 1.2 $ Myrs for the central star, 
while those of Baraffe et al. (1998), give 0.66 $\MSUN$ and
1.6 Myrs, respectively. Both ages are consistent with
that of the rest of the population of Taurus,
$\sim$ 1 - 2 Myrs (Hartmann 2003).

There are two alternatives for the cause of the
high extinction observed towards \coku, either of 
interstellar/intercloud origin, or, 
 if \coku\ were an edge-on disk, of local origin.
However, edge-on disks are underluminous, 
because the direct light from the star gets heavily extinguished,
so the optical spectrum  is stellar 
scattered light (D'Alessio et al. 1999).  
This scattered component
is fainter than the stellar direct flux 
because the scattering region does not cover 4 $\pi$ steradians.
A correction of $A_V$ = 3 applied to this scattered light
component would still leave the star underluminous;
however, with $A_V$ = 3 and its current IR magnitudes, 
the properties of \coku\ agree with those of the
low-mass population of Taurus.
 In addition, the large constrast between 10 and 20 $\mu m$
in \coku\ is much larger than predicted in high inclination
disks with emission in 10 $\mu m$ (Pascucci et al. 2004)

\begin{figure}
\plotone{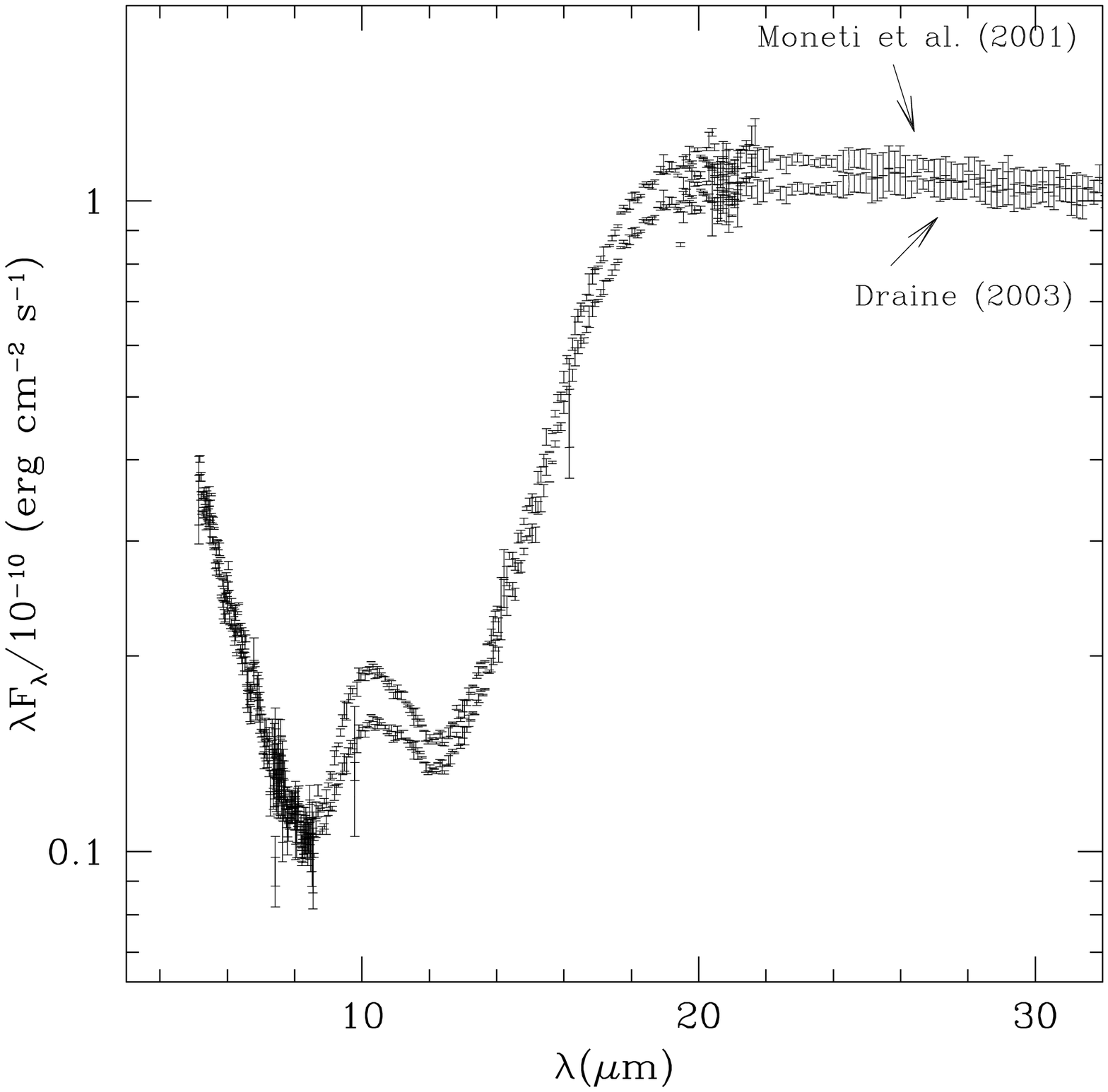}
\caption{Spectrum of Co Ku Tau 4 corrected using 
reddening laws
by Draine (2003) and Moneti et al. (2001)
assuming $A_V$=3. The IRS fluxes are shown by error bars wich represent the
observed error, estimated as 
 half of the absolute
value of the difference in flux from the spectra independently obtained at
the two nod positions of the telescope for each order of Short-Low and
Long-Low.
}
\label{fig_deredd}
\end{figure}

We then adopt the hypothesis that the
extinction towards \coku\ is interstellar/intercloud;
the presence of a scattered light nebulosity $\sim$ 1' from
\coku\ (D. Padgett, unpublished) supports the assumption
that it is located in a high density region of the molecular
cloud. With this assumption,
the IRS spectrum must also be corrected for extinction. 
Figure \ref{fig_deredd} 
shows the IRS spectrum corrected using two different 
reddening laws, one proposed by Draine (2003) and the other one by  
Moneti et al. (2001) 
based on observations from Lutz (1999) and 
10 and 18 $\mu$m profiles from Simpson (1991). 
Between 1.2 and 8 $\mu$m, the reddening law proposed by Moneti et al. (2001) 
agrees, within the uncertainties, with the reddening law 
 recently determined using IRAC data (Indebetouw et al. 2004).
The overall features of the spectrum shown in Forrest \etal (2004) 
are preserved, namely,
a steep, roughly photospheric spectrum shortward of about $8$ $\mu$m, the weak
silicate feature at $10$ $\mu$m, the steep rise between 14 and 20 $\mu$m, and the flattening
of the emission to longer wavelengths.  The shape of the $10$ $\mu$m 
silicate feature is slightly dependent upon the extinction law used.
The flux error at each wavelength was derived by taking half of the absolute
value of the difference in flux from the spectra independently obtained at
the two nod positions of the telescope for each order of Short-Low 
(SL; 5.2-14 $\mu$m, $\lambda/\Delta \lambda \sim 90$) and
Long-Low (LL; 14-38 $\mu$m, $\lambda/\Delta \lambda \sim 90$) 
(See Forrest et al. 2004).

\begin{figure}
\plotone{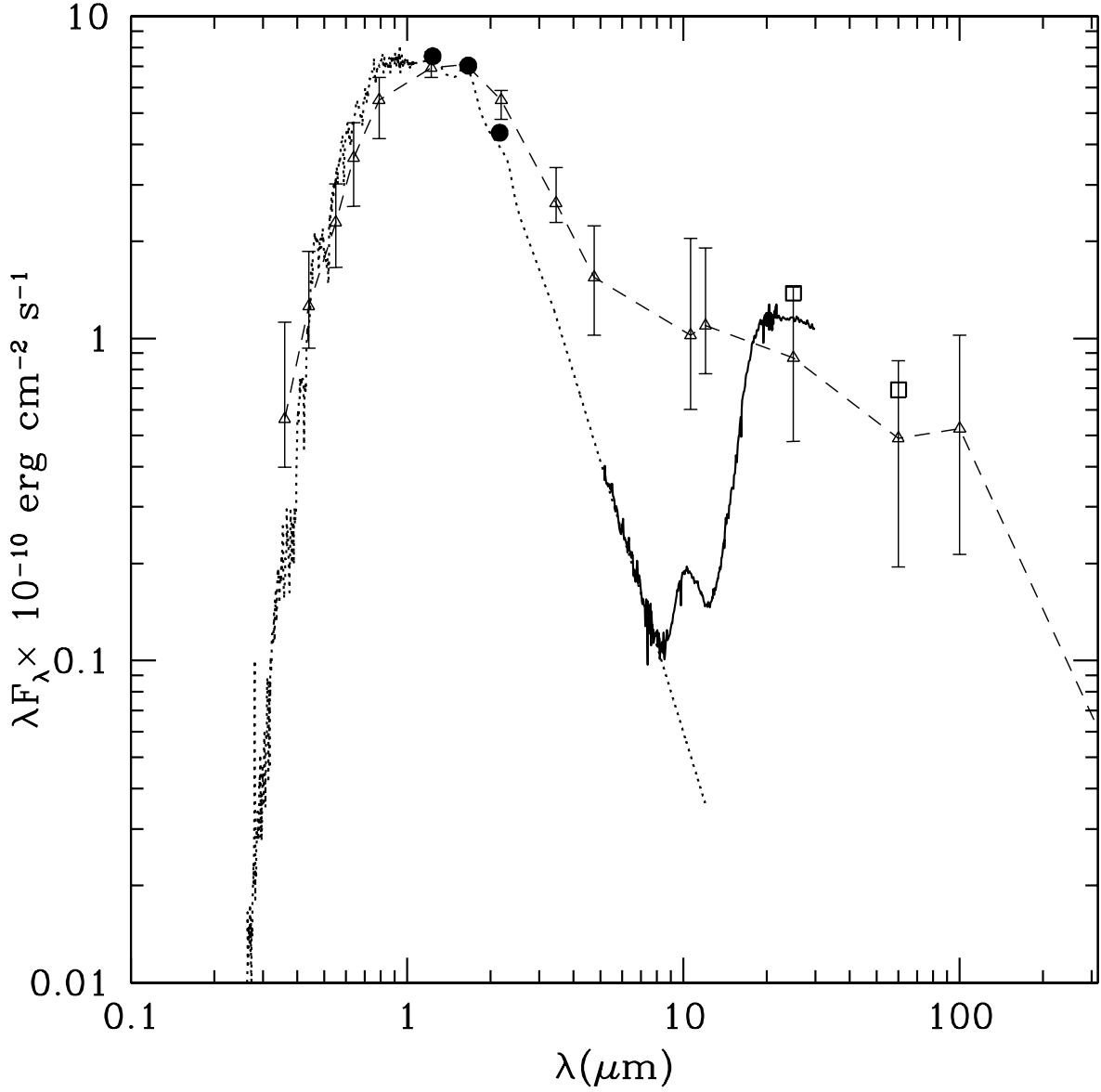}
\vskip1cm
\caption{
Spectral energy distribution of \coku\
(squares from Strom et al. 1989;  
circles from 2MASS; 
IRS portion from Forrest et al. 2004)
corrected using the reddening 
law by Moneti et al. (2001) with $A_V=3$, compared to the median 
SED of Classical T Tauri stars in Taurus  
 (triangles and dashed line). The error bars of the median 
points represent the quartiles (D'Alessio et al. 1999).
We also show a synthetic spectrum for the central star 
taken from  Bruzual \& Charlot (1993), for $T_*=3720$ K and 
$R_*=1.9 \ \RSUN$. 
}
\label{fig_median}
\end{figure}

Figure \ref{fig_median} also shows fluxes from 2MASS and 
long-wavelength measurements
from IRAS from Strom et al. (1989), which may be contaminated 
by extended emission.
For comparison we show a synthetic stellar spectrum from
Bruzual \& Charlot (1993) for $T_*$=3720 K, $R_*=1.9 \ \RSUN$, 
and $M_*=0.5 \ \MSUN$, 
along with the median SED of Classical T Tauri 
stars in the Taurus Association (D'Alessio et al. 1999), 
scaled to the flux in the $\H$ band.
It is evident that the mid infrared emission of 
  CoKu Tau 4 is significantly less than those of 
  typical Taurus stars whose disks lack inner holes. 
In contrast, the flux levels in the 20-35 $\mu$m region are relatively
bright compared with the median Taurus SED.  Our task is to explain
this spectrum.

\section{Model}
\label{secc_model}

We have constructed  models for \coku\ 
that are similar to the one calculated for TW Hya (Calvet et
al. 2002; Uchida et al. 2004).  The basic model for TW Hya is that of
a fairly massive Classical T Tauri disk truncated at an inner radius
of about 4 AU (Calvet \etal 2002); at this radius the optically-thick
disk is frontally-illuminated by the central star.  A major difference
between TW Hya and \coku\  is that the former object exhibits
clear evidence for gas accretion onto the star (Muzerolle \etal 2000;
Batalha \etal 2002).  The SED of TW Hya also exhibits a substantial
excess at 10 $\mu$m and even some small excess emission at wavelengths
as short as $\sim 5$ $\mu$m (Sitko, Lynch, \& Russell 2000).  
Calvet \etal (2002)
interpreted these results to mean that there is a small amount of
optically-thin dust and gas interior to the optically-thick disk wall,
which accounts for the hotter dust emission and the accreting gas.  In
contrast, \coku\  is a WTTS, with a small
H$_\alpha$ equivalent width consistent with stellar chromospheric
emission; the accretion rate
in this system must be extremely small, if present at all 
(Muzerolle \etal 2000).
In addition, the SED of \coku\  is consistent with purely stellar photospheric 
emission at wavelengths $\lambda \lesssim 8$ $\mu$m (Forrest \etal 2004). 
As discussed in the previous section, there are uncertainties at the 
$\sim$~20\%
level in extrapolating the photospheric continuum from shorter wavelengths to
the IRS range; however, the spectral slope at the short-wavelength 
end of the IRS
spectrum is clearly consistent with being purely photospheric, as are 
IRAC colors
of this object between 3.6 and 8 $\mu$m (Hartmann 2004, personal 
communication).

Thus, we model \coku\  as a 
central star with the properties described in \S \ref{secc_star}
surrounded by an optically-thick, truncated disk, with an inner wall at 
a radius $R_{\wall}$. 
Inside this radius the disk is optically thin;
in \S \ref{secc_interior} we estimate the maximum amount of dust in the 
inner hole or 
gap consistent with the spectrum.  Outside  $R_{\wall}$ there is a 
disk which 
might contribute 
to the SED at wavelengths longer than $25$ $\mu$m.  
At $R=R_{\wall}$, the disk receives radiation frontally from the 
central star; as we detail in the following discussion, the wall appears to be
responsible for most, if not all, of the excess emission detected by IRS.

\subsection{Treatment of the inner disk ``wall''}
\label{secc_wall}

\begin{figure}
\plotone{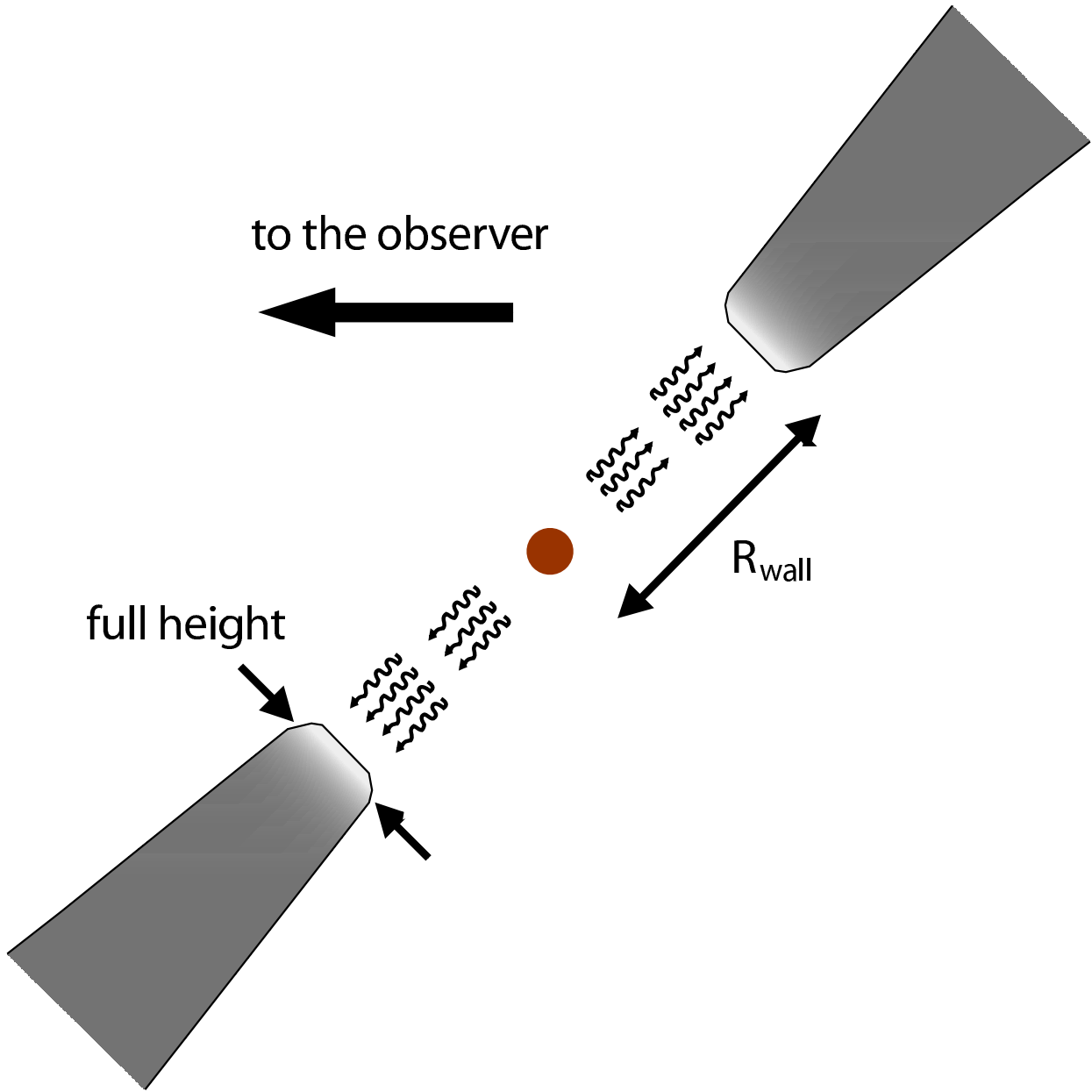}
\caption{Schematic representation of the truncated disk in Co Ku Tau 4. }
\label{fig_wall}
\end{figure}

The wall is the transition between the inner optically thin disk or gap
and the outer disk.  We assume that it is uniform in the 
vertical direction and solve its radial structure. Closer to the star, 
the wall has an optically thin atmosphere, and its radial optical depth 
increases with radius.
Figure \ref{fig_wall} shows schematically the structure of this wall.

The temperature distribution of the wall is calculated using 
the procedure described by Calvet et al. 
(1991, hereafter CPMD; 1992), which 
has been successfully applied to calculate the radial  structure of the wall 
at  the dust sublimation radius in disks around Classical T Tauri Stars 
by Muzerolle et al. (2003).
Since we relax some approximations made in previous papers, 
we write the equations in detail.

We assume that the stellar radiation penetrates the wall 
with an angle zero between the incident direction and the normal 
to the wall's surface, and that the radial thickness of its 
atmosphere is negligible compared to its distance to the central star.
The incident radiation field is separated into two wavelength ranges,
one characteristic of the disk (related quantities
have  subindex ``d'') and one characteristic of the incident
stellar radiation (quantities with subindex ``s'').  
In order to find an analytical solution for the wall
radial distribution of temperature, 
we assume that the opacities are 
constant in the wall atmosphere and that scattering of stellar radiation 
is isotropic.
We also assume that there is no heating source other than the 
incident and scattered stellar radiation.
At every depth into the wall atmosphere the net outward
radial flux at the
disk frequency range is equal to the absorbed radiative flux 
at the stellar frequency range (see CPMD), 
i.e, the Eddington flux ,
$H_d$, is given by

\be
H_d(\tau_s) = {F_0 \over 4 \pi} \alpha [(1+C_1) e^{-\tau_s} +
C_2 e^{- \beta \tau_s} ]
\label{eq_hd}
\en
where $\alpha=1-w$, $w$ is the mean albedo to the stellar
radiation ($w=\sigma_s/\chi_s$), 
$\beta=(3 \alpha)^{1/2}$, $F_0=L_*/4 \pi R_{\wall}^2$ 
and $\tau_s$ is the total mean optical
depth to the stellar radiation in the incident direction (i.e.,
along rays parallel to the disk mid-plane, perpendicular to a
cylindrical surface at $R=R_{\wall}$).
The constants in equation (\ref{eq_hd}) are given by

\be
C_1 = - {3 w \over 1-\beta^2},
\label{eq_c1}
\en

\be
C_2 = {5 w \over \beta [1+(2 \beta/3)](1-\beta^2)}.
\label{eq_c2}
\en

The zeroth moment of the transfer equation can be written as (Mihalas 1978),

\be
{dH_d \over d\tau_d} =  J_d - S_d
\label{eq_dhd}
\en
where $\tau_d$ is the total mean optical depth at the disk frequency range,
$S_d$ is the local source function and $J_d$ is the mean intensity.
CPMD assume strict LTE, i.e., $S_d=B_d=\sigma_R T_d^4/\pi$, 
where $\sigma_R$ 
is the Stefan-Boltzmann constant.
Here, we 
include a term corresponding to the emissivity by scattering
at the disk frequency range, i.e.,
\be
S_d = (\kappa_d B_d + \sigma_d J_d)/(\kappa_d+\sigma_d), 
\en
assuming isotropic scattering (Mihalas 1978).
With this source function,
the zeroth moment of the transfer equation is given by

\be
B_d = J_d - {\chi_d \over \kappa_d} {d H_d \over d \tau_d}.
\label{eq_bd}
\en
where $\chi_d=\kappa_d+\sigma_d$.

Using the Eddington approximation
for the disk radiation field, 
the first moment of the transfer
equation is
\be
{dJ_d \over d \tau_d} = 3 H_d 
\label{eq_djd}
\en
where we have defined $q=\chi_s/\chi_d$. Integrating $J_d$ from
equation (\ref{eq_djd}), using
$J_d(0)=2 H_d(0)$  as the boundary condition, and
substituting it along with $dH_d/d\tau_d$ from equation (\ref{eq_hd})
into equation (\ref{eq_bd}), the temperature as a function of $\tau_d$
can be written as

\be
T_d(\tau_d)^4 = \alpha {F_0 \over 4 \sigma_R } [ C_1^\prime +
C_2^\prime e^{-q \tau_d} +C_3^\prime e^{-\beta q \tau_d}]
\label{eq_temp}
\en
where

\be
C_1^\prime = (1+C_1)\biggl (2+{3 \over q} \biggr )+C_2 \biggl (2+ {3 \over \beta
 q} \biggr )
\en

\be
C_2^\prime = (1+C_1) \biggl( {q \chi_d \over \kappa_d} - {3\over q} \biggr )
\en

\be
C_3^\prime=C_2 \beta \biggl ({q \chi_d \over \kappa_d}  - 
{3 \over q \beta^2} \biggr)
\en

The temperature at the innermost radius of the wall,
at the top of the wall atmosphere 
where $\tau_d= \tau_s=0$,
is $T_0$, 
can be used to define the wall's radius,

\be
R_{\wall} = \biggl [  {\alpha \ L_*  \over 16 \pi \sigma_R}
(1+C_1) \biggl ( 2 + q {\chi_d \over \kappa_d} \biggr ) +C_2 \biggl (2 +
\beta q {\chi_d \over \kappa_d} \biggr )  \biggr ]^{1/2}
{1 \over T_{0}^2}.
\label{eq_rwall}
\en

We use $T_0$ as an input parameter to characterize the atmosphere of the wall.
The value of $R_{\wall}$ 
follows from $T_0$ and the 
adopted dust properties, using equation (\ref{eq_rwall}). 
The mean opacity in the disk wavelength range is calculated using 
the Planck function evaluated at $T_0$ as the weighting function, and 
the mean opacity in the stellar wavelength range is calculated using the 
Planck function at $T_*$. We use Planck means in order to 
have the correct temperature in the optically thin layers of 
the wall atmosphere (Mihalas 1978).

The minimum temperature of the wall is 
obtained when $\tau_d \rightarrow \infty$ and it is given by  
\be
T_{min} =  \biggr [ \alpha {F_0 \over 4 \sigma_R} C_1^\prime \biggl ]^{1/4}  
\label{eq_tmin}
\en

\begin{figure}
\plotone{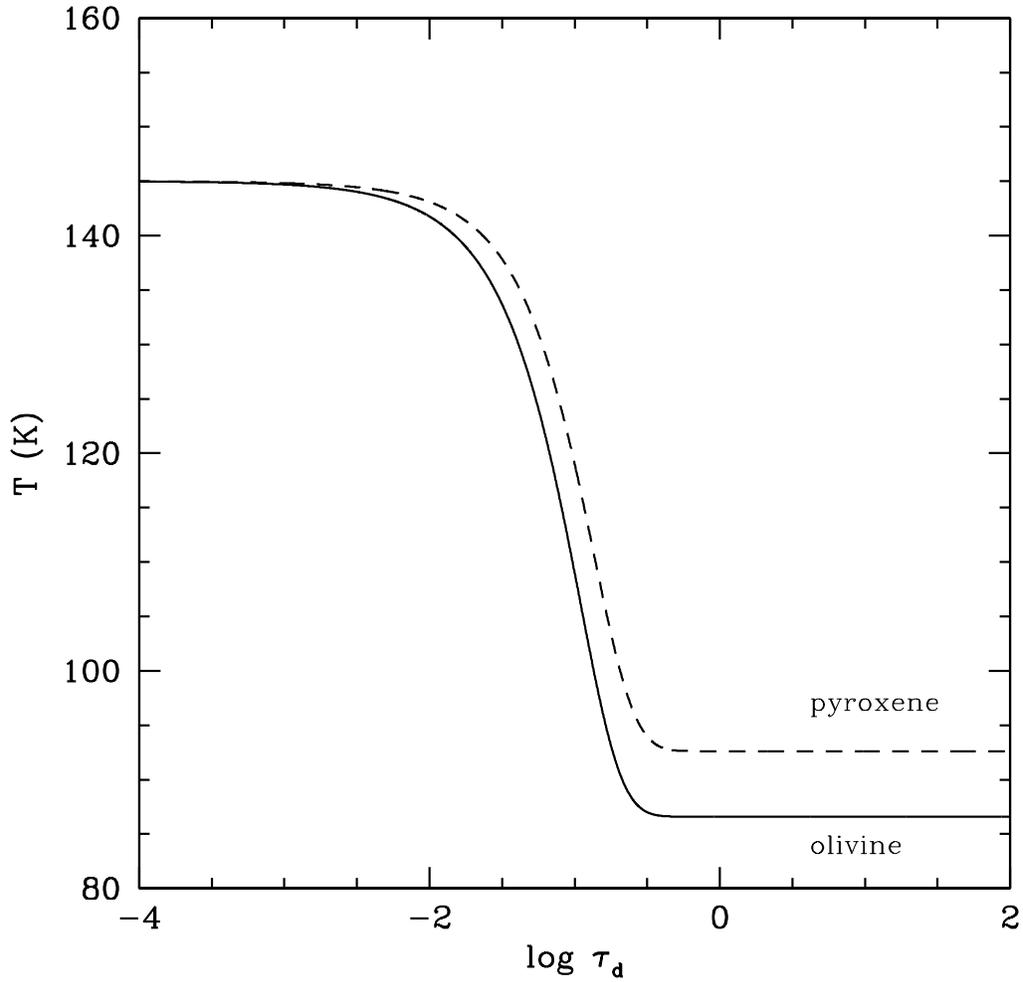}
\caption{Radial distribution of temperature 
for wall models whose synthetic spectrum 
 fits the observed \coku\ spectrum.
Both models have $T_0=$145 K, one contains Mg$_{0.5}$ Fe$_{0.5}$ olivine 
(solid line) and the other one has Mg$_{0.8}$ Fe$_{0.2}$ pyroxene 
(dashed line), to illustrate the effect of the different silicate 
compositions on the temperature distribution.  
}
\label{fig_temp}
\end{figure}

 Figure \ref{fig_temp} shows the temperature as a function of 
the disk mean optical depth (measured from the innermost radius 
of the wall toward larger radii) for  two models  with 
the same value of $T_0$ but different kind of silicate: 
Mg$_{0.5}$ Fe$_{0.5}$ olivine and Mg$_{0.8}$ Fe$_{0.2}$ pyroxene. 
Both models, with appropriated values for the disk inclination angle 
and the wall height, reproduce the observed spectrum of \coku  
(see \S \ref{secc_dust}).
The temperature of the wall as a function 
of optical depth depends on the monochromatic absorption coefficient 
of the dust through the quantity $q$, so it has to be calculated 
for every dust composition attempted to fit the spectrum. 
However, for  a 
given composition, the temperature is independent of the total dust to gas 
mass ratio, since this ratio 
affects the heating and the cooling of 
the grains in the same way, canceling out.

 We have calculated the asymmetry factor $g$  for Mg Fe olivine 
and pyroxene,
the materials that provide the best fit to the observed spectrum 
(see \S \ref{secc_dust}), assuming grains are compact 
spheres and using the method developed by Wiscombe (1979). We 
  note that the asymmetry factor $g$ of these ingredients 
decrease with wavelength. In particular,  
 for a maximum grain size $a_{max}=0.25 \ \mu$m, 
 $g \sim$ 0.6 for $\lambda=0.55 \ \mu$m and $g \sim$ 0.2 
for $\lambda=1.22 \ \mu$m. Taking a Planck
 mean of the $g$ factor 
using the Planck function evaluated at the stellar effective temperature 
as the weighting function,
the mean asymmetry factor for scattering of stellar radiation is 
$<g> \sim $    0.3. This justifies the assumption of isotropic 
scattering ($g \sim$  0) 
we have used to find an analytical solution for the
temperature distribution in the disk.

\subsection{Spectrum of the wall}
\label{secc_spec}

Accounting for the limb brightening produced by the temperature inversion 
in the wall atmosphere, we calculate the emergent intensity 
for each surface element of the wall as seen in the plane of the sky.
The geometry is described in detail in the Appendix. Here we briefly outline 
the required expressions.

 The cosine of the angle 
between the normal to the surface of the wall and the line of sight 
is given by

\be
\cos \Theta = \sin i \  \sin \theta
\label{eq_teta}
\en
where $i$ is the angle between the disk rotation axis and the line of sight
and $\theta$ is such that the coordinate $X$ in the plane of the sky is 
$X=R_{\wall} \cos \theta$.
Each element in the visible surface of the wall 
 has a thermal emergent intensity given by

\be
I_\nu =  {1 \over \cos \Theta} \int_0^\infty B_\nu[T_d(\tau_d)] 
e^{-\tau_\nu/ \cos \Theta} d \tau_\nu
\label{eq_itherm}
\en
where $\tau_\nu = \tau_d (\kappa_\nu/\chi_d)$. The emergent flux is calculated by integrating the emergent 
intensity over solid angle on the visible surface of the wall. The limits 
for the integral are given in the Appendix.

 Using the intensity given by the equation (\ref{eq_itherm})
we obtain a limb brightening effect 
because rays that pierce the limb of the wall (i.e., $x \sim R_{wall}$, see  Figure \ref {fig_geom}) have an optical depth equal unity in layers with 
higher temperature than rays that intersect the wall closer to the center 
(i.e., $x \sim 0$). 
This effect would be  neglected 
if the thermal emergent intensity is
approximated as isotropic and evaluated at $\cos \Theta=1$
with the expression 

\be
I_\nu \approx \int_0^\infty B_\nu[T_d(\tau_d)] e^{-\tau_\nu} d \tau_\nu .
\en

In this approximation, the emergent flux is calculated by multiplying 
the total intensity by the total solid angle subtended by the visible 
side of the wall,
which at a distance $d$ 
from the observer is given by 
\be
\Omega_{\wall}  = 2 \cos i \left( \frac{R_{\wall}}{d} \right)^2
(\delta \sqrt{1-\delta^2}+\sin^{-1} \delta)
\quad {\rm if} \quad \delta < 1
\label{dull1}
\en

\be
\Omega_{\wall} = \pi \left( {R_{\wall} \over d} \right)^2 \cos i
\quad  {\rm if}  \quad  \delta \ge 1
\label{dull2}
\en
where \be
\delta={H_{\wall} \over R_{\wall}} \tan i,
\en 
and $H_{\wall}$ the (vertical) height of the wall photosphere
(measured from the midplane, in the vertical direction). 
Equations (\ref{dull1}) and (\ref{dull2}) are consistent with  the solid angle portion  of the 
expressions used by 
Dullemond, Dominik \& Natta (2001) to calculate  the flux of an isothermal wall.

 Limb brightening effects are important 
when trying to fit a  high resolution  spectrum as 
those of Spitzer-IRS. For instance, we  find that  
taking into account limb brightening the 
temperature $T_0$ required to fit the spectrum is smaller by 5 K   
than the one in the  approximated solution, increasing 
the wall radius by  10 \%. In turn, this translates into 
a height of the  wall 20\% smaller than the one inferred 
for the approximated solution.
Even more important, the 
shape of the 20 $\mu m$ region of the spectrum only can be fitted 
with the detailed model; the approximated model 
shows an emission deficit for all the compositions we have 
tried (described in \S \ref{secc_dust}). 
The 20 $\mu$m flux deficit of the approximated model 
compared to the observed spectrum 
could be wrongly interpreted as
the lack of additional unknown opacity sources.

\subsection{Dust properties}
\label{secc_dust}

We use a mixture of grains composed of silicate
with mass fraction $\zeta_{sil} = $0.0034,
organics with $\zeta_{org}=$0.001, and
troilite with $\zeta_{troi}= 7.68 \times 10^{-4}$.
The grains are assumed to be spheres.
We adopt the standard MRN grain size distribution $n(a) \sim a^{-3.5}$
(Mathis, Rumpl \& Nordsieck  1977), between minimum radius $a_{min}$ and
maximum radius $a_{max}$.
The opacity is calculated using the Mie theory with 
a code developed by Wiscombe (1979).
Optical constants for the organics are taken from Pollack et al. 
(1994, hereafter P94)
and for troilite from P94 and Begemann et al. (1994).
Sublimation temperatures $T_{sil}=1400$ K, $T_{org}=$ 425 K,
and $T_{troi}=$ 680 K are adopted.

The shape of the 10 $\mu$m silicate band and  the position of 
the peak in the spectrum of \coku\  
indicates that the grains responsible for the emission are 
small. We adopt $a_{min}=0.005$ $\mu$m and 
$a_{max}=0.25$ $\mu$m, 
which are consistent with ISM grains (Draine \& Lee 1984). 
If grains were larger than this, then
the peak of the silicate emission would broaden to 
longer wavelengths than observed.

\begin{figure}
\plotone{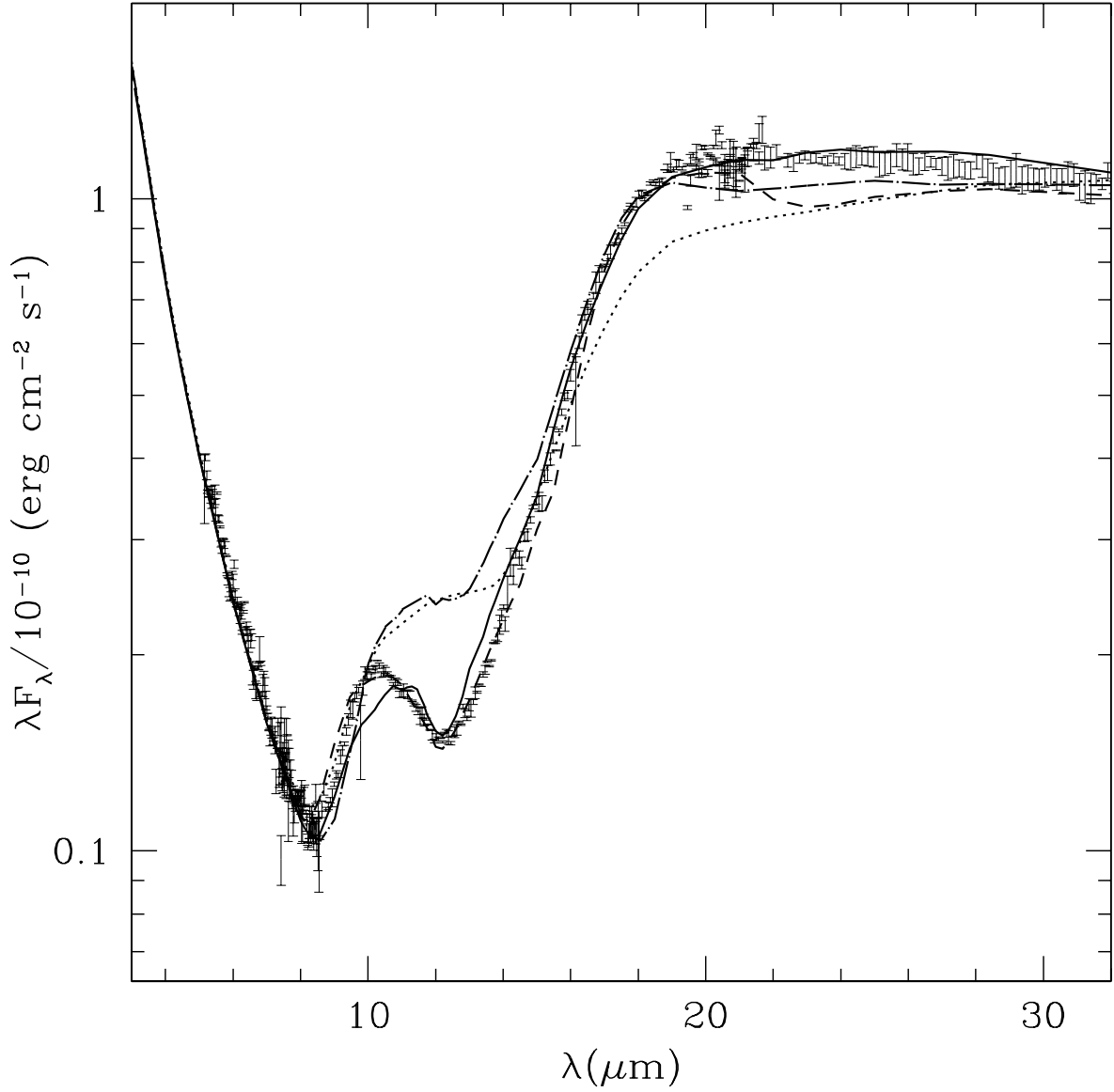}
\caption{Spectra of wall models calculated using  diferent optical constants:
silicates from 
Draine(2003) (dotted line) for $H_{wall}=$280 
$R_*$; glassy silicate of mean cosmic composition 
 Mg$_{0.50}$ Fe$_{0.43}$ Ca$_{0.03}$ Al$_{0.04}$ SiO$_3$ (dashed line)
 with optical 
constants from J\"ager et al. (1994) and  $H_{wall}=$240 
$R_*$ ; glassy bronzite (solid line) with optical constants from 
Dorschner et al. (1988) and $H_{wall}=$ 240 $R_*$; 
oxygen rich circumstellar silicates (dot-dashed line) with optical constants 
from Ossenkopf et al. (1992) and $H_{wall}=$ 320 $R_*$. In all these models 
$T_0=$145 K and $\cos i=$ 0.55. The synthetic spectra are compared to 
the spectrum of \coku\ corrected using Moneti's reddening law. }
\label{fig_sed_draine}
\end{figure}

We have considered several possibilities for the composition
of the silicate (see Henning et al. 
1999\footnote{\tt http://www.astro.uni-jena.de/Laboratory/Database/jpdoc/index.html}).
The observed 10 $\mu$m band is smooth and narrow, with no evidence of 
the substructure characteristic of crystalline components such as  
enstatite, forsterite and silica (e.g., Uchida et al. 2004), 
suggesting the silicate should be glassy and amorphous.
In particular, the shape of the observed 10 $\mu$m band allows 
us to rule out 
glassy bronzite (with optical constants from Dorschner et al. 1988),
crystalline bronzite (optical constants from Henning \& Mutschke 1997), 
crystalline silicate of mean cosmic composition 
 Mg$_{0.50}$ Fe$_{0.43}$ Ca$_{0.03}$ Al$_{0.04}$ SiO$_3$ (optical 
constants from J\"ager et al. 1994), oxygen rich and oxygen deficient 
circumstellar silicate (optical constants from Ossenkopf et al. 1992),
and the absorption coefficient proposed by Draine  
(2003\footnote{\tt http://www.astro.princeton.edu/~draine/dust/dustmix.html} 
 ) 
for interstellar medium dust.
The Draine(2003) mixture and the oxygen rich and deficient 
circumstellar silicate from Ossenkopf et al. (1992) 
produce  an excess flux at the 10 - 13 $\mu$m
wavelength range and a deficit at longer wavelengths compared to the
\coku\ spectrum.
The model spectrum calculated for glassy silicate of mean cosmic composition 
 Mg$_{0.50}$ Fe$_{0.43}$ Ca$_{0.03}$ Al$_{0.04}$ SiO$_3$ (optical 
constants from J\"ager et al. 1994) fits well the 
10 $\mu$m band, but the shape of the spectrum between  
20 and 30 $\mu$m is different from 
the observed one.
Figure \ref{fig_sed_draine} shows some examples of the synthetic 
spectra of models that we have considered do not fit the observed spectrum.
It is important to mention that we have changed the temperature 
$T_0$, inclination angle and height of the wall in the ranges described below
 and we have not found any combination of parameters for these compositions
that produces a good fit to the observed spectrum.

\begin{figure}
\plotone{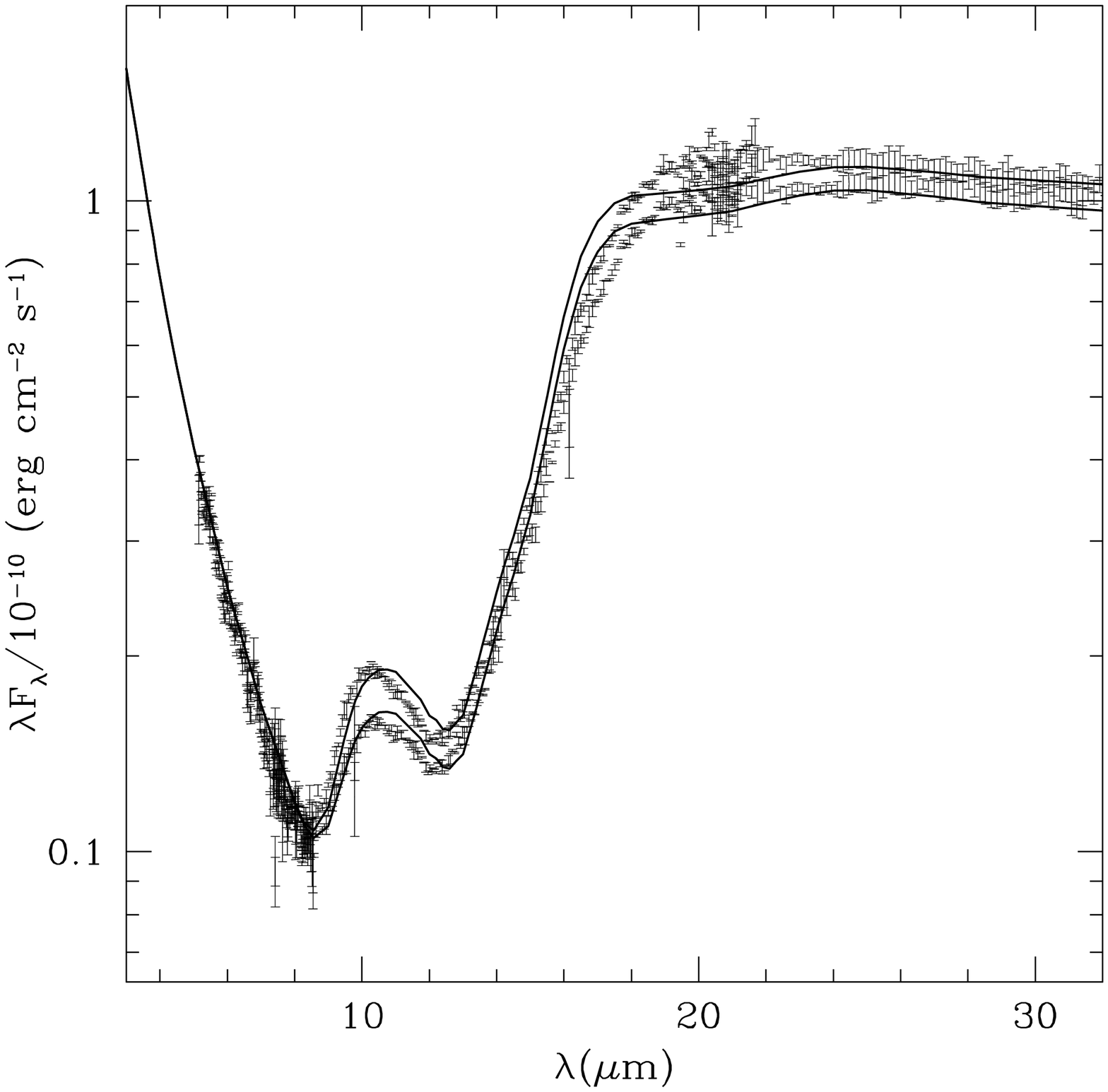}
\caption{Examples of wall model spectra (solid lines) that fit to 
the observed spectrum of \coku\, corrected by the 
reddening laws of Draine (2003) 
(model parameters: $T_0=$ 140 K, $\cos i=$0.45 and 
$H_{wall}=$280 $R_*$) and
Moneti et al. (2001) (model parameters: $T_0=$ 145 K, $\cos i=$0.45 and 
$H_{wall}=$260 $R_*$). The dust 
consists of
small grains ($a_{max}=0.25$ $\mu$m) 
of  glassy olivine, 
with 50 \% Fe and 50 \% Mg, with a small amount of organic and troilite grains 
 (see \S \ref{secc_dust}).  
}
\label{fig_sed1}
\end{figure}

\begin{figure}
\plotone{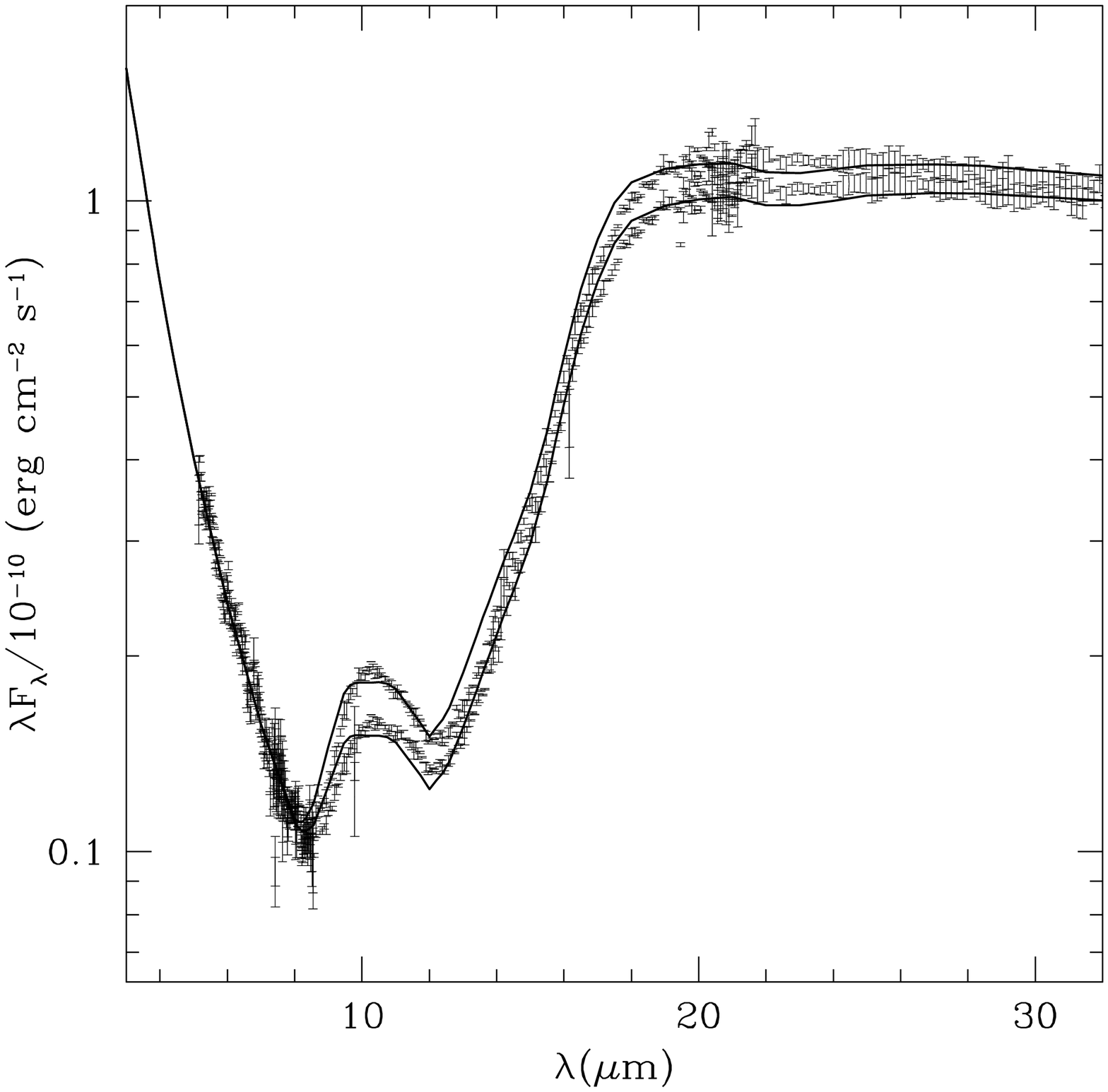}
\caption{Example of wall models spectra (solid lines) that fit to 
the observed spectrum of \coku\, corrected by the 
reddening laws of Draine (2003) 
(model parameters: $T_0=$ 140 K, $\cos i=$0.45 and 
$H_{wall}=$240 $R_*$) and
Moneti et al. (2001) (model parameters: $T_0=$ 145 K, $\cos i=$0.45 and 
$H_{wall}=$240 $R_*$) . The dust 
consists of
small grains ($a_{max}=0.25$ $\mu$m) 
of  glassy pyroxene, 
with 20 \% Fe and 80 \% Mg, with a small amount of organic and troilite grains 
 (see \S \ref{secc_dust}).  
}
\label{fig_sed2}
\end{figure}

On the other hand, amorphous Mg-Fe glassy olivine ( Mg$_{0.5}$ Fe$_{0.5}$ SiO$_4$)
 and glassy pyroxene (in particular, 
Mg$_{0.95}$ Fe$_{0.05}$ SiO$_3$ and Mg$_{0.8}$ Fe$_{0.2}$ SiO$_3$)  
 with optical constants from Dorschner et al. (1995), produce a 
reasonable fit to the spectrum. In particular, the pyroxene has a broader 
$10 \mu$m band than the observed one but fits the 
spectrum for  $\lambda > $ 18 $\mu$m better than the olivine. 
On the other hand, 
the shape of the 10 $\mu$m band predicted for the olivine is 
in better agreement with the observed spectrum.
We find that if olivine or pyroxene has less magnesium then the 
resulting model spectrum has a flatter 10 $\mu$m band compared to the 
observed one.
However, we are not considering 
 both olivine and pyroxene,  
nor the  effect of the grains shape and  porosity, which could affect 
the shape of the bands.
Thus, given the errors and our limited set of 
compositions, we only conclude that both, glassy olivine 
and glassy pyroxene can be present in the wall of CoKu Tau/4, and that 
there is no evidence of a crystalline component.
Figures \ref{fig_sed1} and \ref{fig_sed2}  
show examples of the model spectra calculated for 
each of these ingredients that fit reasonably well 
the observed spectrum. 

For each grain composition we calculate the temperature distribution 
using equation (\ref{eq_temp}) and, in order to fit the observed spectrum, 
we vary the disk inclination angle and the height of the wall, 
which are properties not constrained by other means. 
Thus, for each composition 
we have calculated a small grid of wall models, for $T_0=$130, 135, 140, 145,
 150, 155 and 160 K, $\cos i=$0.25, 0.35, 0.45, 0.55, 0.65 and 0.75, 
$H_{wall} =$ 200, 220, 240, 260, 280 and 300 $R_*$ 
(i.e., 1.77, 1.95, 2.13, 2.3, 2.48 and 2.66 AU). 
Assuming that gas and dust have the same temperature at the wall, 
we can evaluate a characteristic vertical gas scale height as 
$h_{wall} = c_s(T_0)/\Omega_k(R_{wall}) = 1.05 (T_0/140 K)^{1/2} (R/10 AU)^{3/2}$, thus, the wall height we are trying span from 1.5 to 2.5 times $h_{wall}$.
 The values of $H_{wall}/h_{wall}$ are consistent with the  
height where the optical depth to the stellar radiation becomes unity 
times the local gas scale height at $\sim$ 10 AU, in a non-truncated 
accretion disk.

Table \ref{table1} summarizes the parameters that best fit the 
observed spectrum
for each composition and for the two different reddening laws, Draine's and 
Moneti's. 
In this table, {\it olmg50} referes to  Mg$_{0.5}$ Fe$_{0.5}$ SiO$_4$, {\it pyrmg95} to
Mg$_{0.95}$ Fe$_{0.05}$ SiO$_3$ and {\it pyrmg80} to Mg$_{0.8}$ Fe$_{0.2}$ SiO$_3$.
Given the uncertainties in the reddening law, dust composition
 and disk inclination angle, we can conclude that the hole has a radius 
$R_{wall}$ between 9 and 12 AU, or $R_{wall}=10.5 \pm 1.5$ AU, 
corresponding to a maximum temperature in the atmosphere of the wall 
between 140 and 150 K. 
If Draine's reddending law is used or the silicate dust 
is dominated by olivine, the resulting wall is slightly 
colder and farther away than when the Moneti's law is used or 
the dust is dominated by pyroxene.
In each case, a different set of disk 
inclination angles and wall heights are required to fit the 
observed spectrum.
The inclination angle is not well constrained from the comparison between
the synthetic spectrum, however for different compositions, 
inclination angles, temperatures and reddening laws, we have 
a robust estimate of the height of 
the wall relative to its radius,  
$H_{wall}/R_{wall} \approx 0.22 \pm 0.02$.

\begin{table*}
{\small
\begin{center}
\caption[Parameters for the wall]
{Parameters for the wall}
\vspace{2mm}
  \begin{tabular}{|c|cccc|cccc|}
    \hline
\multicolumn{1}{|c}{} & \multicolumn{4}{|c|}{\bf Moneti's law }  & \multicolumn{4}{|c|}{\bf Draine's law }\\\hline
{\bf Silicates }  & { $T_0$  } & { $R_{wall}$} & { $\cos i$}  
& ${ H_{wall}/R_{wall} }$ &  { $T_0$ } & ${ R_{wall} }$ & ${ \cos i}$  & ${ H_{wall}/R_{wall} }$  \\
              &K         &  AU  & & & K & AU & & \\\hline
olmg50	&145	&11.0	&0.35	&0.22/0.24	&140	&12.1	&0.35	&0.20/0.22 \\
	&	&	&0.45	&0.21/0.22	&	&	&0.45	&0.2/0.22\\
	&	&	&0.55	&0.21/0.22	&	&	&0.55	&0.20 \\
	&	&	&0.65	&0.22		&	&	&0.65	&0.20 \\
	&150	&10.1	&0.35	&0.21		&145	&11.0	&0.25	&0.21/0.22 \\
	&	&	&0.45	&0.21		&	&	&0.35	&0.19/0.21 \\
	&	&	&0.55	&0.21		&	&	&	&  \\
\hline
pyrmg95	&140	&10.3	&0.45	&0.24		&140	&10.3	&0.35	&0.22 \\
	&	&	&0.55	&0.24/0.26	&	&	&0.45	&0.21 \\
	&	&	&0.65	&0.24		&	&	&0.55	&0.21/0.22 \\
	&145	&9.4	&0.25	&0.28		&	&	&0.65	&0.21/0.22 \\
	&	&	&0.35	&0.23/0.24	&	&	&0.75	&0.22 \\
	&	&	&0.45	&0.23/0.24	&145	&9.3	&0.25	&0.21 \\
	&	&	&0.55	&0.23		&	&	&0.35	&0.19/0.21 \\
	&	&	&0.65	&0.23		&	&	&0.45	&0.19 \\
	&150	&8.6	&0.25	&0.23/0.25	&	&	&	&	\\
	&	&	&0.35	&0.21	&	&	&	&	\\
	&	&	&0.45	&0.21	&	&	&	&	\\
\hline
pyrmg80 &140	&10.4	&0.35	&0.25	&135	&11.4	&0.45	&0.23 \\	
	&	&	&0.45	&0.24	&	&	&0.55	&0.23 \\
	&	&	&0.55	&0.24	&140	&10.4	&0.25	&0.24/0.25 \\
	&	&	&0.65	&0.25	&	&	&0.35	&0.22 \\
	&145	&9.4	&0.35	&0.23/0.24 &	&	&0.45	&0.20/0.22 \\
	&	&	&0.45	&0.23	&	&	&0.55	&0.20 \\
	&	&	&0.55	&0.23	&	&	&0.65	&0.22 \\
	&150	&8.7	&0.35	&0.20/0.22 &	&	&0.75	&0.22 \\
	&	&	&	&	&145	&9.4	&0.25	&0.23 \\
	&	&	&	&	&	&	&0.35	&0.19/0.21 \\
	&	&	&	&	&	&	&0.45	&0.19 \\
	&	&	&	&	&	&	&0.55	&0.19 \\
\hline
\end{tabular}

\label{table1}
\end{center}
}
\end{table*}

\subsection{Limits on dust in the inner disk}
\label{secc_interior}

In this section, we quantify an upper limit of the dust  mass in the
inner hole of \coku.
The dust mean temperature is calculated assuming radiative equilibrium 
between the dust grains and the stellar radiation field. 
We assume the inner hole, filled with gas and small grains, 
has a constant optical depth at 10 $\mu$m (measured from the midplane 
to the surface of the disk) as a function of radius. 
 The mass in dust in the inner disk 
is given by

\be
M_{sil}^{hole} \approx  6.6 
\left( {\tau_{10\mu\mrm{m}} \over \kappa_{10\mu\mrm{m}}} \right) 
\left( {\zeta_{sil} \over 0.0034} \right)
\left( {R_{\wall} \over 10 \ AU} \right)^2 {\rm lunar \ \ masses}, 
\en
For olivine Mg$_{0.5}$ Fe$_{0.5}$ SiO$_4$, with a $a_{max}=$0.25 $\mu$m, 
 the absorption coefficient  at the 10 $\mu$m peak is 
$\kappa_{10\mu\mrm{m}}=9$ cm$^2$ g$^{-1}$, assuming 
a silicate dust to gas mass 
ratio $\zeta_{sil}=0.0034$. Thus, for $\tau_{10\mu \mrm{m}}$=0.01, and a wall 
radius $R_{wall}\sim $ 10 AU,  the mass 
in silicates in the inner hole is 0.007 lunar masses.

\begin{figure}
\plotone{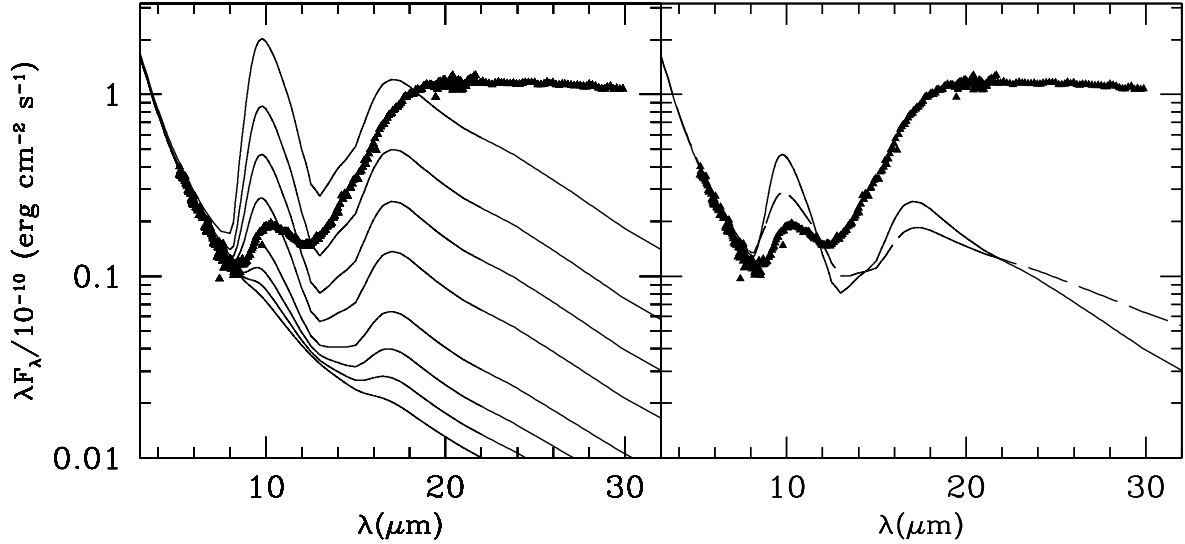}
\caption{Spectrum of dust in the inner disk compared to the 
IRS spectrum of Co Ku Tau 4 corrected using the reddening law by
Moneti et al. 2001.  
The adopted dust is composed by glassy olivine, 
with 50 \% Fe and 50 \% Mg. Left panel: 
for $a_{max}=0.25$ $\mu$m and each curve corresponds to a different 
total silicate mass $M_{sil}^{hole}=$ 0.00015, 0.0004, 0.0007, 
0.0015, 0.004, 0.007, 0.015 and  0.04 
lunar masses (corresponding to $\tau_{10\mu \mrm{m}}=$0.0002, 0.0005, 
0.001, 0.002, 0.005, 0.01, 0.02, 0.05, from bottom to top). 
Right panel: for $\tau_{10\mu\mrm{m}}=$ 0.01; each curve corresponds 
to a different maximum grain size $a_{max}=$ 0.25  (solid line), 
 10 (long-dashed line) 
(corresponding to $M_{sil}^{hole}=$ 
0.007 and 0.013 lunar masses, respectively). Biger maximum grain
sizes, for the same optical depth at 10 $\mu$m, produce a similar
spectrum than the case for $a_{max}=10 \ \mu$ m, but for 
a different total dust mass. For instance, $a_{max}=$ 1 cm corresponds 
to 0.36 lunar masses, $a_{max}=$ 1 m, to 3.6 lunar masses. In each of these 
cases,
$a_{min}=0.005 \ \mu$m and the exponent of the size distribution is $p=3.5$.
}
\label{fig_interior}
\end{figure}

The left panel of
Figure \ref{fig_interior} shows the contribution to the 
SED of the dust 
in the hole (assuming a sublimation temperature $T_{sil}=1400$ K)  
 for different values $\tau_{10\mu \mrm{m}}$.
The temperature of the hottest dust grains inside this inner hole 
is such that the emissivity is larger at 10 $\mu$m than at 20 $\mu$m, 
in contrast to the observed spectrum. The only way to compensate for this 
effect is to assume a particular radial dependence of $\tau_{10\mu \mrm{m}}$, 
for instance, if $\tau_{10\mu \mrm{m}}$ is very small 
in the hole  and increases rapidly 
at  $R \sim R_{\wall}$ (which is precisely how the wall 
is defined).
For these small grains, we find that 
the upper limit in the mass of silicates in 
the inner disk is $M_{sil}^{hole} <$ 0.0007 lunar masses. A higher mass than 
this limit would distort the 10 $\mu$m silicate band.

The right panel of
Figure \ref{fig_interior} 
illustrates  the effect of increasing 
the maximum  grain size. 
 For a fixed value of the optical depth at 
10 $\mu$m, $\tau_{10\mu \mrm{m}}=$0.01, we show the spectrum of 
the interior for different maximum grain sizes. Since the absorption 
coefficient at 10 $\mu$m decreases with  $a_{max}$, each spectrum 
corresponds to a different silicate mass. 
If grains are larger, the mass limit increases. For instance, if the 
grains in the inner hole have $a_{max}=$ 10 $\mu$m, then 
$M_{sil}^{hole} <$ 0.0013 lunar masses.

There are not enough constraints on the properties of the whole 
disk, but  speculating 
that before the gap formation it was a typical 
 $\alpha$-irradiated disk with 
$\Mdot=10^{-8}$ $\MSUNYR$ and  $\alpha=0.01$, 
the mass in gas of the unperturbed
inner region (for $R < 10$ AU) is $M_{gas}=3 \times 10^{-3}$ $\MSUN$. 
For the adopted silicate dust to gas mass ratio, one would expect to find 
$M_{sil}^{hole} = 250$ lunar masses in the hole.
The lost mass in grains could be locked in 
larger bodies or could have been accreted by the star.
In the first case, bodies would have grown to $\sim 30$ km 
to account for this mass, assuming that the same power law holds even for this
huge size.

\subsection{The outer disk contribution}
\label{secc_disk}
 Given the uncertainty of the flux at 60 $\mu$m
and that there is only an upper limit for the flux
at 1.3 mm ($F_\nu < $ 15 mJy, Osterloh \& Beckwith 1995), 
it is difficult to constrain the outer disk. 

The wall treatment assumes plane-parallel geometry.  We therefore 
 estimate that the heated wall region has a radial thickness 
 $\Delta R \lesssim H_{\wall}$, where $H_{\wall}$ is the thickness of the 
 wall photosphere.   Then, measured either radially or vertically, 
 \be 
 \tau_d \sim \kappa_d \Sigma\,. 
 \en 
 If we estimate $\kappa_d \sim 2$  cm$^2$ g$^{-1}$ (see \S \ref{secc_dust}), 
then  
 \be 
 \tau_d \sim \Sigma (\mrm{cgs}). 
 \en 
 Our disk models with typical parameters for T Tauri stars 
 ($\Mdot \sim 10^{-8}$ $\MSUNYR$, $\alpha = 0.01$) yield 
 $\Sigma \sim 10$ g cm$^{-2}$ at 10 AU, so $\tau_d \sim 10$, 
 consistent with our optically-thick wall treatment but low enough 
 not to contribute significantly at 1.3 mm (the mass 
of the wall 
 is only $\sim 10^{-4}$ $\MSUN$).  
Of course, the disk 
 structure might be quite different, with a pile-up of material 
 at the disk wall (a ring); in this case we can limit the amount 
 of mass present, depending upon the adopted opacity, assuming that 
 the wall 
and the disk 
are optically thin, taking the Rayleigh-Jeans limit, 
 and 
taking the minimum temperature 
of the wall, $T_{min} \sim $ 85 K (equation \ref{eq_tmin})
as the maximum temperature of the disk, 
 \be 
\left( {M_{disk} \over 0.001 \MSUN} \right) <  
\left( { F_{\nu} \over 15 \mrm{mJy}}\right)
\left({ \kappa_{1.3\mrm{mm}} \over 0.01 \mrm{cm}^2 \ \mrm{g}^{-1}} \right)^{-1} 
\left({ T_{min} \over 85 \ \mrm{K}} \right)^{-1}
 \en 
where the adopted opacity corresponds to a maximum grain size $\sim$ 1 mm.
Smaller or bigger maximum grain sizes would imply a higher mass 
given the same flux. 
To better constrain the outer disk, it would be very useful to measure the fluxes at 70 $\mu$m and longer wavelengths.

\section{Origin of the hole in CoKu Tau 4}

What is the origin of the central clearing in the \coku\ disk?  
Processes such as grain coagulation and disk accretion are expected to
play a role in preferentially reducing the opacity of the inner disk
relative to the outer disk due to both the high surface densities and
shorter orbital periods in this region.  So, as alluded to earlier (cf.
section 3.4), a possible explanation for the observed SED is that grain
growth has led to the production of bodies up to ~30 km in size,
rendering the inner disk optically thin.  Whether or not models of disk
accretion and grain growth can reproduce in detail the observed SED of
\coku\ is an interesting topic for future study.

In the context of the model presented here, we show that the observed
SED of \coku\  is consistent with a dramatic decrement in the dust
opacity at a radial distance of 10 AU. An interesting interpretation of
this  structure is that the growth of large bodies has reached an
advanced state, producing a planet with a mass large enough to open a
substantial gap, allow the interior material to accrete onto the star,
and prevent through orbital resonances the progress toward the star of
disk material further out.
This effect has been suggested as the reason for
central clearings in disks around T Tauri stars (e.g., Marsh \& Mahoney 1992, 1993) and debris disks (Jura et al. 1993; Backman, Gillett \& Low 1986). We
consider this also to be the most probable origin for the clearing in 
\coku; in fact, the case for planet formation is somewhat stronger than in
the older systems, after considering several alternatives as follows.

{\bf Radiation pressure or Poynting-Robertson effect}. 
In an optically thin disk
of this size, the lifetime of dust grains 10 AU from a star like \coku\
 due to these effects would be about $10^{5}$ years (see, e.g. Chen
\& Jura 2001, or Burns, Lamy \& Soter 1979). In a disk optically thick
to most of the starlight, the time scales are substantially longer ($>>$ 1
Myr), as the material would have to be removed an optical depth at a time.
The outer disk of \coku\ is quite optically thick even in the mid
infrared; so are the inner disks around most other Classical T Tauri stars.
Thus these effects are very unlikely to be important here. Even if they
were, they would not naturally lead to a very sharp edge.

{\bf Sharp change in gravitational field at 10 AU} (see Duschl 1989).  
This
effect, which has been invoked to explain the central clearings in
galactic nuclei, could be used here if the disk were self-gravitating at
$r >$ 10 AU; the inner part would fall in and be accreted by the star on the
viscous time scale (about $10^{5}$ years; Quillen et al. 2004). However,
the limit on the mass of the \coku\ disk from millimeter-wave
observations (Osterloh and Beckwith 1995), $10^{-3}~M_\odot$, is too
small for the disk to be self-gravitating.

{\bf Radial grain segregation}. 
In debris disks around A stars young enough for
some gas still to be present, it has been predicted that the combined
effects of radiation pressure and gas drag would lead to central clearings
in the distribution of small grains (Takeuchi and Artymowicz 2001). The
radiative requirements are similar to those of radiation pressure and
Poynting-Robertson drag, though; in the present case the disk is too
optically thick, and the star too cool, for this process to produce a
central clearing within 1 Myr.

{\bf Stellar companion}. 
Were \coku\ to have a stellar or planetary companion, a clear gap
would be produced on timescales shorter than the 1 Myr stellar age,
if the accretion timescale were sufficiently short, and if the companion
were sufficiently massive both to open a gap in the disk (i.e., if its
Hill radius were to exceed the original half-density thickness of the
disk), and to prevent migration. Accretion of the inner-disk material
would take place on the viscous timescale of $10^{5}$ years, which is
indeed sufficiently short. A terrestrial-size planet would be unable to
produce a gap or prevent migration. A stellar companion could do both. No
such companion has yet been observed. The visible and infrared properties
of \coku\ are accurately consistent with a single M1.5
pre-main-sequence star; a companion star is constrained to have luminosity
of at most a small fraction of that of the M1.5 star, for which the
luminosity is already small on the scale of stars. Thus it is not
impossible that the central clearing has been produced and maintained by a
very low-mass star which has so far escaped detection. 
The range of stellar masses for this to 
    work is so narrow, however, that this seems 
    less likely an option than a substellar  
    companion.

A companion in a wide range of masses in the giant planet or brown dwarf
regime would produce such an empty, 
sharp-edged clearing  as we see. Upon formation
of the companion, material further out than the strongest orbital
resonances would be held up by frequent collisions at these resonances 
 until the mass held up were similar to the mass
of the companion. Material inside these resonances would simply be
accreted on the viscous time scale. Quillen et al. (2004) have recently
simulated the dynamics of a \coku-like disk that undergoes
giant-planet formation, and found that clearings like the one 
observed were
produced for companions as small as 0.1 Jupiter masses. 

Observations can be used to test the scenario proposed in 
this paper to interpret the spectrum of \coku. 
For instance, molecular line observations
could probe whether the outer disk is gaseous; continuum millimeter
observations can be used to quantify the outer disk mass; 
spatially resolved images could confirm that there is a sharp 
decline in the dust column density around 10 AU and might reveal 
whether the disk has the non-axisymmetric structure that is expected to be
induced by the planet (e.g., Quillen et al. 2004).

\section{Discussion and Summary}

The spectrum of \coku\  is consistent with that  expected from a disk 
truncated at an inner radius $\sim $ 10 AU. No detectable material is 
in the inner hole; the spectrum indicates at most 
0.0007 lunar masses of small silicate grains in the inner hole region.
Even though \coku\ is much younger than TW Hya, the hole in its inner disk
is larger  and apparently more evacuated of dust and gas (judged
by the lack of near-infrared dust emission and gaseous accretion),
and its outer disk appears to be much less massive.
This comparison clearly indicates that age is not the sole parameter
in determining disk evolution.  That other parameters influence
the properties of pre-main sequence disks has been clear for some
time.  For example, nearly half the stars in Taurus lack inner disks,
some (but not necessarily all) of which can be explained by the
presence of perturbing binary companions (e.g., Mathieu \etal 2000,
and references therein).

Because it seems likely that even binary systems
once possessed accretion disks (see, e.g., Bate, Bonnell, \& Bromm 2003),
there must be a wide range of evolutionary timescales for protoplanetary
disks.  The reasons for this range of timescales 
are not clear, but
we conjecture that initial angular momentum could play a crucial role;
disks of similar mass but much smaller initial radii are more likely
to fragment and to  coagulate faster (given higher initial disk surface
densities).   This conjecture predicts that the outer radius of the
\coku\ disk is small; observational efforts to explore this
possibility should be made.  In any case, given the low disk mass estimate,
\coku\ may be a system observed just barely before the initial disk
material disappears.

Finally, to summarize our results, we find that
an inner disk wall at $R \sim $ 10 AU with a half-height $H_{\wall} \sim$ 2 AU
, is responsible for the observed  
excess between 8 and 30 
$\mu$m. This inner wall has an optically thin atmosphere, closer to the star, 
with a temperature decreasing with radius. 
The highest temperature of the wall 
is $\sim 145$ K, but decreases rapidly to $\sim 85$ K in the deepest layers.
Both the 10 and 20 $\mu$m bands emerge from the optically thin hotter layers.
The spectrum for $\lambda > 20$ $\mu$m emerges from the optically thick
layers of the wall. 
At even longer wavelengths the whole wall becomes optically thin, and 
it is expected that the outer disk dominates the SED.
Using an upper flux limit at 1.3 mm (Osterloh \& Beckwith 1995), and
assuming that the  wall is so optically thin at this wavelength that
it  does not contribute 
to the flux,  we estimate the mass of the outer disk to be less than
0.001 $\MSUN$ for a maximum grain size of 1 mm.
The central star dominates the spectrum at $\lambda < 8$ $\mu$m.

From an analysis of the 10 $\mu$m feature, we find that most 
of the dust in the wall should be glassy Fe Mg olivine  and/or pyroxene,
with no evidence of a crystalline component. We believe this represents the initial, unprocessed state of the dust in
the disk that formed around \coku\ . 

\section{Appendix}

The coordinate system $x$, $y$ and $z$ is centered on the star, 
the $z$ axis 
is the disk rotation axis, and the plane $(x,y)$ is the disk midplane.
We use also cylindrical coordinates  
$R,\theta$ and $z$. The wall 
is a cylinder with radius $R_{\wall}$ and total height $2 H_{\wall}$.
Points on the wall surface have coordinates  $z, x=R_d \cos \theta$ and
 $y=R_d \sin \theta$.
We use another coordinate system 
$X$,  $Y$,  $Z$, also centered on the star, 
where $Z$ is along the line of sight and $(X,Y)$ is the 
plane of the sky. Both systems coincide when the disk is pole-on.
The transformation between coordinate systems 
is

\be
x=X
\en
\be
y=Y \cos i - Z \sin i
\en
\be
z=Y \sin i + Z \cos i
\en

or

\be
X=x
\en
\be
Y=y \cos i + z \sin i
\en
\be
Z=z \cos i - y \sin i
\en

The normal to the cylinder farthest wall (which is the one the observer 
will see) is

\be
\hat{n} = -\hat R = - \cos \theta \hat{x} - \sin \theta \hat{y}
\en
thus, the cosine between the line of sight 

\be
\hat{Z} = -\sin i \  \hat{y} + \cos i \  \hat{z},
\en
and the normal to the visible surface of the wall is
\be
\cos \Theta = \hat{Z} \cdot  \hat{n} = \sin i \  \sin \theta
\en

\begin{figure}
\plotone{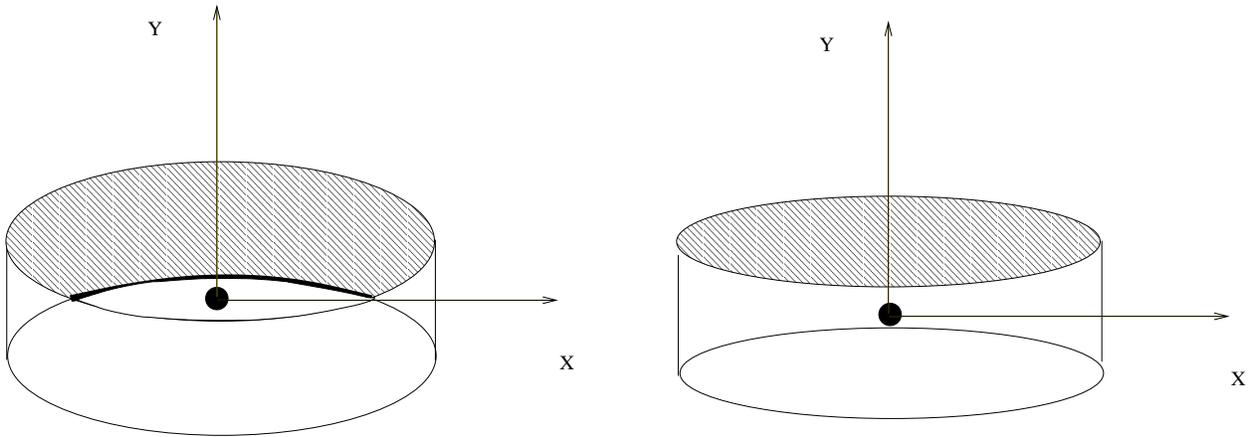}
\caption{Schematic representation of the visible surface of 
the wall (dashed surface)  as seen by the observer for two inclination angles,
one corresponding to $\delta < 1$ (the star is visible) and 
the other one to $\delta >1$ (the star is invisible). 
}
\label{fig_geom}
\end{figure}

A surface element of the visible area is 
$dA = dX \ dY = - R_d \ \sin \theta \  d \theta \ \ dY$.
To find the 
limits of the visible surface, we have to define 
two ellipses (given by the projections of the upper and lower tops
of the wall; see Figure \ref{fig_geom}). For the upper ellipse

\be
{Y_{up} \over R_d} = \pm \cos i \sqrt{1 - X^2/R_d^2} + {H \over R_d} \sin i
\en

For the lower ellipse,

\be
{Y_{down} \over R_d} = \pm \cos i \sqrt{1 - X^2/R_d^2} - {H \over R_d} \sin i
\en

Which can be written also in terms of $\theta$, using $X=x=R_d \cos \theta$,

\be
{Y_{up} \over R_d} = \pm \cos i \sin \theta + {H \over R_d} \sin i
\en

\be
{Y_{down} \over R_d} = \pm \cos i \sin \theta - {H \over R_d} \sin i
\en

Both ellipses intersect at critical angles: $\theta_c$ and $\pi-\theta_c$, where
$\theta_c$ is given by

\be
-\cos i \sin \theta_c + {H \over R_d} \sin i = \cos i \sin \theta_c - {H \over R_d} \sin i
\en
\be
\sin \theta_c =  {H \over R_d} \tan i = \delta
\en

Depending on the inclination angle $i$, both ellipses can intersect if $\delta < 1$ or not
if $\delta > 1$ (see Figure \ref{fig_geom}).
If $\delta > 1$, then the integration region is defined by the upper ellipse
and $Y_{up}^- < Y < Y_{up}^+$, $0 < \theta < \pi$.
If $\delta < 1$,  there are two regions: if $0 < \theta < \sin^{-1} \delta$ 
then $Y_{up}^- < Y < Y_{up}^+$ but if $ \sin^{-1} \delta  < \theta  < \pi/2$ 
then $Y_{up}^- < Y < Y_{down}^+$.

\acknowledgments This work is based on observations
made with the Spitzer Space Telescope, which is operated
by the Jet Propulsion Laboratory, California Institute of
Technology under NASA contract 1407. Support for this work
was provided by NASA through Contract Number 1257184 issued
by JPL/Caltech and through the Spitzer Fellowship Program,
under award 011 808-001.
PD and RFH acknowledge grants from CONACyT and DGAPA.
NC and LH acknowledge support by NSF 
grant AST-9987367 and NASA grant NAG5-10545.
FK akcnowledges support provided by NASA through the Spitzer 
Felloship Program, under award 011 808-001.

\end{document}